\documentclass[twocolumn,preprintnumbers,superscriptaddress,altaffilletter,amsmath,amssymb]{revtex4}

\usepackage{epsfig,amsfonts,amstext,afterpage}

\newcommand{\tr}[1]{\operatorname{tr}\left[#1\right]}

\begin{document}
\preprint{SI-HEP-2020-30, P3H-20-072}

\title{Gauge anomalies in the Standard-Model Effective Field Theory}

\author{Oscar Cat\`a}
\email{oscar.cata@uni-siegen.de}

\author{Wolfgang Kilian}
\email{kilian@physik.uni-siegen.de}

\author{Nils Kreher}
\email{nils.kreher@uni-siegen.de}

\address{\it%
  University of Siegen, Department of Physics,
  D--57068 Siegen, Germany
}

\date{\today}

\begin{abstract}
If the Standard Model is understood as the first term of an effective
field theory, the anomaly-cancellation conditions have to be worked
out and fulfilled order by order in the effective field-theory
expansion. We bring attention to this issue and study in detail a
subset of the anomalies of the effective field theories at the
electroweak scale. The end result is a set of sum rules for the
operator coefficients. These conditions, which are necessary for the
internal consistency of the theory, lead to a number of
phenomenological consequences when implemented in analyses of
experimental data. In particular, they not only decrease the number of
free parameters in different physical processes but have the potential
to relate processes with different flavor content. Conversely, a
violation of these conditions would necessarily imply the existence of
undetected non-decoupling new
  physics associated with the electroweak energy scale.
\end{abstract}

\maketitle

{\emph{Introduction}}. Anomalies in quantum field theories occur
whenever a classical symmetry does not survive the process of
quantization. Such effects are actually not uncommon and are typically
accompanied by profound phenomenological consequences. In the strong
interactions, the chiral anomaly explains the $\pi^0\to\gamma \gamma$
decay \cite{Bell:1969ts}, indirectly confirms the number of quark
colors, and it also implies the existence of pions as Goldstone bosons
\cite{tHooft:1979rat}. The axial $U(1)_A$ anomaly, in turn, justifies
the mass range of the $\eta^\prime(958)$ boson
\cite{tHooft:1986ooh}. All of these effects are associated with
anomalous global symmetries. Local symmetries have a different status,
and the existence of gauge anomalies has more severe
consequences for a relativistic quantum field theory, namely 
the absence of a unitary S matrix.

If the theory at hand is assumed to be complete, gauge anomalies deem
it inconsistent. Instead, if we are dealing with an effective field
theory (EFT) \cite{Appelquist:1974tg,Weinberg:1978kz}, the presence of
an anomaly can be amended with the addition of a Wess-Zumino 
term, which is associated with heavy integrated-out degrees of
freedom. A paradigmatic  
example is the Standard Model (SM) with the top quark integrated out. In
this theory, the  
top quark contribution to the anomaly is retained by a Wess-Zumino
term~\cite{DHoker:1984mif} which
cancels the anomaly associated with the active fermions. The term vanishes 
again in the deep infrared regime, where all fermions are integrated out.
In general, EFTs
with an anomalous gauge symmetry indicate the existence of
non-decoupling physics beyond the SM
and must be augmented either by 
non-local interactions, by extra degrees of freedom, or by
non-polynomial terms of Wess-Zumino type~\cite{Ball:1988xg,Preskill:1990fr}.

The SM is a formally complete theory, in particular free
from gauge anomalies. However, strong indications suggest
that it should be considered as
the first term of an EFT expansion. An EFT provides a well-defined
low-energy approximation to a field theory, organizing amplitudes and
observables as a power-series expansion in terms of an inverse heavy
mass scale $\Lambda$, beyond which the theory is no longer applicable.

The Standard-Model Effective Theory
(SMEFT) \cite{Buchmuller:1985jz,Hagiwara:1993ck,Grzadkowski:2010es}
extends the Lagrangian of the SM by
gauge-invariant local operators ${\cal O}_j^{(D)}$ of dimension $D>4$,
built from its 
elementary fields. Formally, 
\begin{align}
  \label{eq:EFT series}
  {\cal{L}}_{\textsc{SMEFT}}
  =
  {\cal{L}}_{\textsc{SM}}
  + \sum_{D>4}^{\infty} \frac{1}{\Lambda^{D-4}}
  \sum_{j}^{n_D} c_j^{(D)} {\cal{O}}_j^{(D)}
\end{align} 
where the sum over $j$ is finite, i.e., each order of this (infinite)
dimensional expansion contains a finite number of operators. The SMEFT
Lagrangian and the associated perturbative expansion at each order
systematically encode the potential effects of unresolved new degrees
of freedom beyond the SM, even without knowledge about the actual
high-energy theory. The framework is valid if there is a clean
separation between the new-physics scale $\Lambda$ and the electroweak
scale~$v$, assuming that the limit $v/\Lambda\to 0$ is well-behaved
within perturbation theory.  The number of independent operators of
SMEFT for each dimension $D$ can be determined either by explicit
construction \cite{Grzadkowski:2010es} or by formal methods
\cite{Henning:2015alf}. By assumption, the theory has no extra
infrared degrees of freedom, and non-local and Wess-Zumino terms are
also absent.

In the SM,
anomaly cancellation holds separately for each fermion generation due
to the hypercharge assignments. However, there is no proof that these
assignments are enough to render the full SMEFT
anomaly-free. Actually, extending the arguments 
of Ref.~\cite{Preskill:1990fr} to the $1/\Lambda^2$ sector, it is
logically possible to develop gauge anomalies at NLO in the EFT
expansion, while having an anomaly-free SM. Accordingly, in order to
have an EFT free of anomalies, one should explicitly 
check that they cancel at each order of the EFT expansion. 

The SMEFT, commonly truncated to contain only dimension-six operators,
has been extensively used in data analysis at the
LHC~\cite{deFlorian:2016spz}, and further for estimating the physics
potential of future
experiments~\cite{Cepeda:2019klc,deBlas:2018mhx,Dainese:2019rgk,dEnterria:2016sca,Mangano:2017tke,Barklow:2017suo}. However,
to our knowledge there is no proof of perturbative unitarity of its
gauge sector. Rather, applications rely on an implicit assumption that
the properties of the SM carry over without modification. As we will
show explicitly below for the dimension-six operators, this assumption
is not warranted.  At the quantum level, the SMEFT generates its own
gauge anomalies at each order in $1/\Lambda$. These are local
contributions to amplitudes that violate Ward identities, and thus
unitarity, which cannot be removed by renormalization.  As a
consequence, anomaly cancellation can only be achieved through the
enforcement of a set of sum rules that relate
the different Wilson coefficients.  The number of
independent parameters is therefore reduced in a model-independent way
by a consistency requirement. If the SMEFT is the low-energy limit of
an anomaly-free UV model, such sum rules are necessarily satisfied. In
a generic bottom-up analysis, they must be imposed by hand as extra
conditions on the SMEFT parameter space.
\begin{figure}
  \begin{displaymath}
    \includegraphics{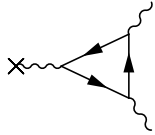}
  \end{displaymath}
  \caption{Feynman graphs contributing to the anomaly in the SM.
    The external lines correspond to any
    combination of the SM gauge bosons. The cross
    indicates a momentum vector inserted in place of the polarization vector.
    All anomalies cancel for each complete SM fermion generation.}
  \label{fig:A-SM} 
\end{figure}
\begin{figure}
  \begin{gather*}
    \includegraphics{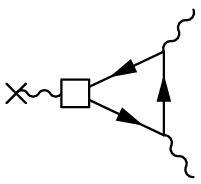}
    \qquad
    \includegraphics{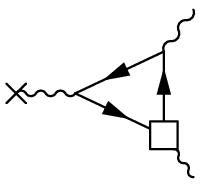}
    \qquad
    \includegraphics{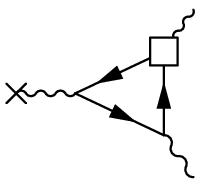}
  \end{gather*}
  \caption{Fermion-loop graphs contributing to the anomaly in the SMEFT at
    dimension~6, with conventions as in Fig.\ref{fig:A-SM}.
    The square indicates a vertex correction that originates
    from a dimension-6 SMEFT operator.}
  \label{fig:A-SMEFT-chiral} 
\end{figure}
\vskip 0.2cm
{\emph{Anomalies in EFTs}}. The problem of identifying the classes of
operators that lead to an anomaly can be systematically solved by
using well-known methods in algebraic renormalization (for a review,
see e.g., \cite{Piguet:1995er}). The renormalization of gauge
theories requires the introduction of a gauge-fixing condition and
ghosts terms. After their introduction, the Lagrangian possesses a
BRST symmetry, which translates into a set of Slavnov-Taylor
identities (STI) for the correlators of the theory. The STI ensure
unitarity of each loop order $n$ of the perturbative effective action
$\Gamma^{(n)}$ through the non-linear functional equation
$\mathcal{S}\Gamma^{(n)}=0$, where $\mathcal{S}$ is the Slavnov-Taylor
differential operator. Explicit expressions for the SMEFT case have 
been given in~\cite{Corbett:2019cwl}. 

In general, this condition is not automatically fulfilled at the
quantum level. Instead, the general solution is
\begin{equation}\label{eq:QAP}
  \mathcal{S}\Gamma^{(n)}= \int d^4x\,\sum_j Q_j,
\end{equation}
where $Q_j$ are a set of local operators. The STI implies
$Q_j=0$. Introducing the anticommuting space-time and BRST
differentials $d$ and $s$, respectively, each of the offending terms
can be removed by renormalization if the cohomology is trivial, i.e.,
all the $Q_j$ are equivalent to a BRST variation up to total
derivatives, $Q_j=d P_j^{(1)} + s P_j^{(2)}$, where $P_j^{(k)}$ are
local operators. Otherwise, the STI are not satisfied, which signals
the existence of an anomaly. As an algebraic structure, the anomaly is
the combined cohomology of $d$ and $s$ in the space of local operators
with fixed operator dimension and unit ghost number. In general, this
space is non-empty and there exist local operators which potentially
violate the Slavnov-Taylor identities, unless their coefficients
vanish.

For standard renormalizable gauge theories, including the SM, this
problem was solved long ago. For each single gauge group, there is at
most one contribution to the anomaly, namely
$Q=\varepsilon_{\mu\nu\lambda\rho} X^{\mu\nu}X^{\lambda\rho}x$, where
$X^{\mu\nu}$ is the field strength, and $x$ denotes the associated
ghost field.  These are the famous ABJ triangle anomalies
\cite{Adler:1969gk,Bell:1969ts}, which were initially discovered by
the computation of the graph in Fig.~\ref{fig:A-SM}.  In the SM, there
are several of these diagrams, depending on the combinations of
external gauge fields, which can only vanish through cancellations of
the different fermion loops. The vanishing of all triangles can be
written as a set of sum rules involving the different quantum numbers
of fermions. All the sum rules are satisfied in the SM, hence the
anomaly vanishes and a Fock space can be defined.

\begin{table*}
\begin{ruledtabular}
\begin{tabular}{lcc}
  Triangle & SMEFT & EWChL\\ \hline
  $\langle G G B\rangle$
           & $2c_{\varphi q}^{(1)} - c_{\varphi u} - c_{\varphi d} = 0$
                   & $2c_{V 1} - c_{V 4} - c_{V 5} = 0$
  \\
  $\langle W W B\rangle$
           & $6y_\varphi c_{\varphi q}^{(1)}
             + 12y_qc_{\varphi q}^{(3)} + 2y_\varphi c_{\varphi l}^{(1)}
             + 4y_lc_{\varphi l}^{(3)} = 0$
                   & $12y_q \mathrm{Re}\,[c_{V 3}]
                     + 4y_l \mathrm{Re}\,[c_{V 9}] - 3c_{V 1} - c_{V 7} = 0$
  \\
  $\langle B B B\rangle$
           & ${\mathcal{A}}\equiv 6y_q^2c_{\varphi q}^{(1)}
             + 2y_l^2c_{\varphi l}^{(1)} - 3y_u^2c_{\varphi u}
             - 3y_d^2c_{\varphi d} - y_e^2c_{\varphi e} = 0$
                   & ${\mathcal{B}}\equiv 6y_q^2c_{V 1}
                     + 2y_l^2c_{V 7} - 3y_u^2c_{V 4} - 3y_d^2c_{V 5}
                     - y_e^2c_{V 10} = 0$
  \\
  $\langle B B W_3\rangle$
           & $(12y_q c_{\varphi q}^{(3)}
             + 4y_l c_{\varphi l}^{(3)})y_{\varphi} + {\mathcal{A}} = 0$
                   & $3y_qc_{V 2} + y_lc_{V 8} - {\mathcal{B}} = 0$
  \\
  $\langle W_3 W_3 B\rangle$
           &
                   & $6y_qc_{V 2} + 2y_lc_{V 8} - 3c_{V 1} - c_{V 7} = 0$
  \\
  \hline
  $\langle RR B\rangle$
           & $6c_{\varphi q}^{(1)} + 2c_{\varphi l}^{(1)}
             - 3c_{\varphi u} - 3c_{\varphi d} - c_{\varphi e} = 0$
                   & $6c_{V 1} + 2c_{V 7} - 3c_{V 4} - 3c_{V 5} - c_{V 10} = 0$
  \\
\end{tabular}
\end{ruledtabular}
\caption{\label{tab:table1} Sum rules enforcing anomaly cancellation
  for the different triangle diagrams in both the SMEFT and the
  EWChL.}
\end{table*}

The same procedure can be applied to the SMEFT, where now the
effective action is a series expansion in inverse powers of
$\Lambda$. The STI should be fulfilled at each order in the EFT
expansion. The anomaly structures can a priori be more general than
the SM ones and, as a result, one expects that the conditions that
ensure anomaly-cancellation in the SM are enlarged when the NLO terms
in SMEFT are considered.

The formalism to solve the cohomology problem for gauge theories with
operators of arbitrary dimension is known~\cite{Barnich:1994mt}. If
the gauge group contains $U(1)$ factors, at each order in the EFT
expansion there are potential anomaly contributions beyond the
ABJ-type one. A basis for the dimension-six anomaly is provided by
operators of the form $Q^{(6)}=R^{(6)}b$, where $R^{(6)}$ are
operators of the dimension-six SMEFT Lagrangian, and $b$ is the
$U(1)_Y$ (hypercharge) ghost.  \vskip 0.2cm {\emph{Triangle anomalies
    in SMEFT}}. In this letter we will concentrate on the subset of
parity-odd operators without fermion fields, namely
\begin{align}
(\varphi^\dagger \varphi)
  \varepsilon^{\mu\nu\rho\sigma}\tr{G_{\mu\nu}G_{\rho\sigma}}b&\\
(\varphi^\dagger \varphi)
  \varepsilon^{\mu\nu\rho\sigma}B_{\mu\nu}B_{\rho\sigma}b&\\
(\varphi^\dagger \varphi)
  \varepsilon^{\mu\nu\rho\sigma}\tr{W_{\mu\nu}W_{\rho\sigma}}b&\\
(\varphi^\dagger \varphi)
  \varepsilon^{\mu\nu\rho\sigma}R_{\mu\nu\alpha\beta}R_{\rho\sigma}^{\,\,\,\,\,\,\,\alpha\beta}b&\\
(\varphi^\dagger \tau^a\varphi)
  \varepsilon^{\mu\nu\rho\sigma}W_{\mu\nu}^aB_{\rho\sigma}b&. 
  \label{eq:A-SM}
\end{align}
where $G_{\mu\nu}$, $W_{\mu\nu}$ and $B_{\rho\sigma}$ are the field
strength tensors of $SU(3)$, $SU(2)$ and $U(1)_Y$, respectively,
$\varphi$ is the Higgs field, $R_{\mu\nu\alpha\beta}$ is the Riemann
tensor, and $\tr{}$ denotes the trace over gauge indices. In the
broken phase, where the Higgs field is expanded
around its vacuum expectation value, the coefficients of these
operators can be computed from the triangle graphs
of Fig.~\ref{fig:A-SMEFT-chiral}, where
each diagram has a single dimension-six insertion. The integrands
become entirely analogous to the SM case, hence no new explicit loop
calculation is necessary.  At NLO in SMEFT, one can also build
diagrams with three external gauge bosons and scalar loops, but they
can be shown not to contribute to the anomalies.

In order to evaluate the diagrams, one needs to identify the NLO
corrections to the gauge boson couplings to fermions. The presence of
dimension-six operators corrects the SM fields and parameters, which
have to be brought back to canonical form (see
e.g., \cite{Brivio:2017btx}). Once this is done, one finds that the
gauge interactions to fermions have flavor- and
family-dependent corrections, which come
from the operators~\cite{Grzadkowski:2010es}
\begin{align}
  \label{eq:operatorsi}
  \left[Q_{\varphi l}^{(1)}\right]_{ij}
  &=(i\varphi^\dagger\!\!\stackrel{\leftrightarrow}{D}_{\!\mu}\!\varphi)(\bar{l}_i\gamma^\mu l_j)
  \\ 
  \left[Q_{\varphi e}\right]_{ij}
  &=(i\varphi^\dagger\!\!\stackrel{\leftrightarrow}{D}_{\!\mu}\!\varphi)(\bar{e}_i\gamma^\mu e_j)
  \\
  \left[Q_{\varphi l}^{(3)}\right]_{ij}
  &=(i\varphi^\dagger\!\!\stackrel{\leftrightarrow}{D}_{\!\mu}\!\varphi^{\!\!\!\!\!\!\!\!\!a}\,\,\,\,)(\bar{l}_i\tau^a\gamma^\mu l_j)
  \\
  \left[Q_{\varphi u}\right]_{ij}
  &=(i\varphi^\dagger\!\!\stackrel{\leftrightarrow}{D}_{\!\mu}\!\varphi)(\bar{u}_i\gamma^\mu u_j)
  \\
  \left[Q_{\varphi q}^{(1)}\right]_{ij}
  &=(i\varphi^\dagger\!\!\stackrel{\leftrightarrow}{D}_{\!\mu}\!\varphi)(\bar{q}_i\gamma^\mu q_j)
  \\
  \left[Q_{\varphi d}\right]_{ij}
  &=(i\varphi^\dagger\!\!\stackrel{\leftrightarrow}{D}_{\!\mu}\!\varphi)(\bar{d}_i\gamma^\mu d_j)
  \\
  \left[Q_{\varphi q}^{(3)}\right]_{ij}
  &=(i\varphi^\dagger\!\!\stackrel{\leftrightarrow}{D}_{\!\mu}\!\varphi^{\!\!\!\!\!\!\!\!\!a}\,\,\,\,)(\bar{q}_i\tau^a\gamma^\mu q_j)
  \\
  \left[Q_{\varphi ud}\right]_{ij}
  &=(i\tilde{\varphi}^\dagger D_{\!\mu}\varphi)(\bar{u}_i\gamma^\mu d_j)
\label{eq:operatorsf}
\end{align}
where $\tau^a$, $a=1,2,3$ are the Pauli matrices.  The fermion
generation indices are explicitly shown. There are 
also universal corrections, associated with the oblique parameters $S$
and $T$ through the operators
$Q_{\varphi \Box}=(\varphi^\dagger \varphi)\Box(\varphi^\dagger
\varphi)$,
$Q_{\varphi D}=(\varphi^\dagger D^\mu \varphi)^*(\varphi^\dagger D_\mu
\varphi)$ and
$Q_{\varphi WB}=(\varphi^\dagger \tau^a
\varphi)W^a_{\mu\nu}B^{\mu\nu}$. However, these contributions to the
triangles cancel 
automatically once all the fermions are considered, precisely because
of their universal character.  The operator $Q_{\varphi ud}$
does not contribute to the triangles, since one cannot build any
diagram with it. The structure of the
corrections coming from the remaining operators in
eqs.~(\ref{eq:operatorsi})-(\ref{eq:operatorsf}) is SM-like, except
for the neutral $W_3^\mu$ gauge boson, which has interactions not
proportional to the identity and to right-handed fermions.
 
We will perform our analysis in the gauge basis, where generation
mixing is only present in the Yukawa terms. Since the anomalies under
consideration are independent of the fermion masses, in this basis
the anomaly conditions are generation-diagonal. 
The resulting sum rules are listed in Table \ref{tab:table1}, where
$R$ stands for the gravitational field. Each of
the SMEFT coefficients contains the sum over all the generations,
e.g., $c_{\varphi u}=\sum_j c_{\varphi u}^{(jj)}$. The first four
conditions correspond to the same triangle topologies as in the SM,
whereas the last two are new configurations of SMEFT at NLO. The
remaining triangles either trivially vanish or reproduce one of the
previous conditions. For instance, the triangles
$\langle G G W_3\rangle$, $\langle W_3 W_3 W_3\rangle$ or
$\langle RR W_3\rangle$ are nonzero but do not provide additional
constraints.

The sum rules above imply that there are just two independent
parameters, which we may denote as $c_{\varphi f}^{(3)}$ and
$c_{\varphi f}^{(1)}$. The solution in terms of them is
\begin{align}
  \label{eq:c3}
  c_{\varphi f}^{(3)}
  &\equiv
    c_{\varphi_q}^{(3)} = c_{\varphi l}^{(3)},
  \\
  \label{eq:c1}
  c_{\varphi f}^{(1)}
  &\equiv
    \frac{1}{y_q} c_{\varphi q}^{(1)}
    = \frac{1}{y_u} c_{\varphi u}
    = \frac{1}{y_d} c_{\varphi d}
    = \frac{1}{y_l} c_{\varphi l}^{(1)}
    = \frac{1}{y_e} c_{\varphi e}.
\end{align}
Anomaly cancellation requires that the trace (over flavor) of the NLO
SMEFT coefficients $c_{\varphi j}$ containing singlet $SU(2)$ fermion
currents have to scale as the ratio of the hypercharges. In
turn, the coefficients of the triplet fermion currents for leptons
and quarks have the same trace. 

Notice that these conditions open up a very rich spectrum of
(anomaly-free) flavor phenomenology. They also entail that fits with
generation-universal Wilson coefficients have substantially less free
parameters than normally assumed. 

We have performed a number of checks of the previous conditions by
exploring the matching to SMEFT of some anomaly-free models with
an additional $U(1)$ gauge group. In all cases, we have found that the
conditions are fulfilled. An independent cross-check also comes from
the renormalization evolution of the different coefficients. The
anomaly-cancellation sum rules that we have found provide constraints
that must be respected in any valid renormalization scheme and for any
renormalization scale~$\mu$. We took the running of the relevant
coefficients from~\cite{Jenkins:2013zja}, which list the complete
one-loop evolution of SMEFT at NLO without imposing anomaly
cancellation. The first thing to notice is that the
coefficients in
eqs.~(\ref{eq:c3}) and (\ref{eq:c1}) are not closed under
renormalization, but involve other
operators. We have checked that by keeping only the operators of
eqs.~(\ref{eq:operatorsi}--\ref{eq:operatorsf})
and neglecting fermion masses, the
constraints in eqs.~(\ref{eq:c3}) and (\ref{eq:c1}) are indeed
invariant under the renormalization-group flow. With these simplifications
in place, this part of the renormalization-group equations reduces to
\begin{align}
  \mu\frac{d}{d\mu} c_{\varphi f}^{(3)}
  &= \frac{7}{3}g_2^2  c_{\varphi f}^{(3)},
  &
    \mu\frac{d}{d\mu} c_{\varphi f}^{(1)}
  &= \frac{41}{3} g_1^2c_{\varphi f}^{(1)}.
\end{align}
However, the renormalization-group evolution of the
coefficients~(\ref{eq:c3}--\ref{eq:c1}) contains, in the general case,
contributions from four-fermion and
Higgs operators, and indirectly mix with the whole NLO SMEFT
Lagrangian. Since anomaly cancellation requires invariance under the
renormalization-group flow for the whole set of operators, the
corresponding coefficients must be subject to additional sum rules
that are not associated with the graphs of
Fig.~\ref{fig:A-SMEFT-chiral}. The determination of these sum rules is
beyond the scope of the present paper.

So far we have concentrated our attention on the SMEFT. However, at
the electroweak scale one can build another EFT, which goes under the
name of the Higgs-Electroweak Chiral Lagrangian (HEW$\chi$L) 
or Higgs Effective Field Theory (HEFT)~\cite{Feruglio:1992wf,Contino:2010mh}. 
This
EFT is linked to scenarios of new-physics which are strongly-coupled
at the electroweak scale (see e.g., the discussion in
\cite{Buchalla:2016bse,Cohen:2020xca}). As opposed to SMEFT, the
leading-order is not the SM, but a more general framework where the
Higgs is not necessarily a $SU(2)$ doublet. However, when it comes to
the anomaly, the arguments spelled out above carry over
unchanged. As with the SMEFT, we will be concerned with the
anomalies that can be computed from the diagrams of
Fig.\ref{fig:A-SMEFT-chiral}. The corrections to the gauge
interactions of fermions contain universal and flavor-specific
components \cite{Buchalla:2013wpa,Cata:2013sva}.  As before, only the
flavor-specific corrections are relevant.  In the basis of
Ref.~\cite{Buchalla:2012qq}, there are 10 NLO operators that belong to
this class. The resulting conditions for the Wilson coefficients, in
the notation of Ref.~\cite{Cata:2015lta}, are summarized in Table
\ref{tab:table1} where, as before, each coefficient implicitly
contains the trace over generation indices. We note that the HEW$\chi$L has
one additional sum rule, which is associated with the anomaly
\begin{align}
  \varepsilon^{\mu\nu\rho\sigma}
  \tr{U^\dagger W_{\mu\nu} U \tau_3}\tr{U^\dagger W_{\rho\sigma}U \tau_3}b,
\end{align}
whose SMEFT counterpart is a dimension-eight operator. The matrix $U$ is a 
(non-linear) function of the electroweak Goldstone modes.
The solutions for the NLO coefficients read
\begin{align}
  \label{eq:c0nl}
  c_{0}
  &\equiv
    c_{V 8} = c_{V 2},
  \\
	\label{eq:c1nl}
	c_{1}
	&\equiv
   {\mathrm{Re}}\,[c_{V 3}]={\mathrm{Re}}\,[c_{V 9}]
	\\
  \label{eq:c2nl}
  c_{2}
  &\equiv    
	    \frac{1}{y_q} c_{V 1}
    = \frac{1}{y_u} c_{V 4}
    = \frac{1}{y_d} c_{V 5}
    = \frac{1}{y_l} c_{V 7}
    = \frac{1}{y_e} c_{V 10}.
\end{align}
The one-loop renormalization of the HEW$\chi$L has been worked out in
\cite{Buchalla:2017jlu,Alonso:2017tdy,Buchalla:2020kdh}. As a
cross-check of the previous conditions, we have verified that, when
the fermion masses are neglected, the relations above are
renormalization-group invariant. The reason in this case is that the
flavor-specific part of the beta functions for each coefficient is
proportional to the hypercharges of the corresponding fermions.

\vskip 0.2cm 
{\emph{Conclusions}}.
The conditions that render the SM anomaly-free do not
automatically carry over to the EFTs at the electroweak scale, namely
the SMEFT and the HEW$\chi$L. Rather, the cancellation of anomalies has to
be implemented order by order in the EFT expansion.
In the present work we have concentrated on a
subset of NLO anomalies, namely those that can be computed with gauge
boson triangle diagrams. This results in a series of sum rules which
impose constraints on the gauge-fermion interactions at NLO. 

It is interesting to remark that anomaly operators
at NLO can also contain matter fields. By
inspection, one finds dimension-six operator structures of 
the generic form
\begin{equation}
  \tilde M^{(1)}_{\mu\nu}B^{\rho\sigma}b,
  \quad
  \tr{\tilde M^{(3)}_{\mu\nu}W^{\rho\sigma}}b,
  \quad
  \tr{\tilde M^{(8)}_{\mu\nu}G^{\rho\sigma}}b,
  \label{eq:A-BSM}
\end{equation}
where $\tilde M^{(1,3,8)}_{\mu\nu}$ are parity-odd
antisymmetric tensors of dimension four built from matter fields. The 
superscripts denote the representation of the tensor under
the relevant SM gauge group. CP-violating phases should also be taken
into account.

The sum rules presented in this paper are universally valid, i.e.,
they are the infrared manifestation of anomaly-free extensions of the
SM and guarantee the consistency of the EFT. Practical advantages are
the reduction of parameters in phenomenological analyses,
e.g., in global fits to LHC data, or correlations between processes
with charged lepton currents and the analogous processes with charged
quark currents, e.g., using the condition
$c_{\varphi q}^{(3)}=c_{\varphi l}^{(3)}$. Interestingly, failure to
satisfy the sum rules would be a clear indication of the existence of
non-decoupling physics that cannot be described within the
current EFT formulations, i.e., new degrees of freedom
below the TeV scale. The analysis 
of the full set of anomaly operators at NLO and the associated sum
rules 
will be addressed in a future publication~\cite{CKK20}.

\vskip 0.2cm
{\emph{Acknowledgments}}. We thank Thorsten Ohl and Tilman Plehn
for useful discussions.
This research was supported by the Deutsche Forschungsgemeinschaft
(DFG, German Research Foundation) under grant 396021762 -- TRR 257.

\bibliographystyle{unsrt}

\end{document}